\documentstyle[11pt,paspconf]{article}

\begin{document}

\title{New Observing Techniques}
\author{D.L. Welch}
\affil{Department of Physics \& Astronomy, McMaster University \\
       1280 Main St W, Hamilton, ON L8S 4M1 Canada}

\begin{abstract}
The influence of new techniques on the discovery and characterization
of pulsating variables has been enormous. In this paper, I will
review the methods and results of a number of research programmes
which have dramatically altered our ability to study variable stars
and will likely bear fruit for many years into the future. Specifically,
I will touch on results from the MACHO and EROS Projects, the HIPPARCOS
mission, "flux difference" photometry and a handful of new algorithmic 
advances which I consider to be important.
\end{abstract}


\section{Introduction}

The pace at which new and better instrumentation has become available
to astronomers has accelerated in the last decade. Furthermore, 
improvements in algorithms and computers have allowed the astronomer
to embark on "computationally expensive" projects which were unthinkable
only ten years ago. In this review, I select a handful of research
programmes which have had a large impact on the study of variable stars
and review several relatively new techniques which I feel will lead to
improvements in analysis in the near-future. I have made no attempt to
be complete as completeness may be next to godliness, but it is also
next to impossible in the time and space permitted. I apologize for my
(many) errors of omission.

\section{Massive Photometry}
\subsection{MACHO}
\begin{center}
{\bf http://wwwmacho.mcmaster.ca/}
\end{center}
The MACHO Project is the first of the microlensing surveys I will discuss.
A joint collaboration of the US Department of Energy (Lawrence Livermore
National Laboratory), National Science Foundation (Center for Particle
Astrophysics), and the Australian Commonwealth Science and Industrial
Research Organization (Australian National University/Mount Stromlo 
Observatory), the Project's primary goal is to determine the characteristics
of non-luminous or under-luminous matter in the Milky Way by detecting
the amplification of starlight as a result of gravitational microlensing.

\newpage

The state of the MACHO Project during 1996, including a description of 
equipment and computational procedures, is given by Cook {\it et al.} 1997.
Briefly, the dedicated 50-inch telescope at Mount Stromlo Observatory,
uses a custom-built camera containing two arrays of four 2048$\times$2048
CCD detectors to obtain simultaneous red and blue images of field in the
Large Magellanic Cloud (LMC), Small Magellanic Cloud (SMC), and the bulge
of the Milky Way galaxy. Since 1992, over 50,000 image pairs have been
obtained in the course of approximately 1800 nights. The photometry is
processed by a multi-CPU system which ensures that data obtained on one
night is completely processed before the next night. At present, some
of our fields have as many as 1500 different observation epochs.

The suitability of such a database for variable star work is obvious and
there have been many variable star papers published, primarily discussing
LMC results. I will briefly describe each of the major results:

\begin{itemize}
\item {\sl Double-mode RR Lyrae} --- Alcock {\it et al.} (1997a) reported
the discovery of 75 RRd stars in the LMC, all new. Pulsation parallaxes
for the ensemble suggest an LMC distance modulus of 18.48 $\pm$ 0.19 mag
for the LMC.

\item {\sl Double-mode Cepheids} --- Alcock {\it et al.} (1995, 1998a, 1998b), 
Welch {\it et al.} (1997), and Rorabeck (1997) have reported the detection
of large numbers of double-mode Cepheids in the LMC and SMC. These detections
reveal the presence of first and second overtone beating in addition to stars
beating in the fundamental and first overtone stars. Furthermore, the period
ratios and location of stars in the instability strip are quite clearly
correlated with metallicity in the manner described by Morgan \& Welch (1997).
Singly-periodic second overtone Cepheids have also been identified
in the LMC.

\item {\sl Classical Cepheids} --- Welch {\it et al.} (1997) has reported
the progress on singly-periodic classical Cepheid variables. The complete 
sample of 1477 LMC Cepheids has been analysed and mean lightcurve 
properties have been derived. Detectable amplitude changes are see in 
about 1\% of the sample.

\item {\sl R CrB Variables} --- Alcock {\it et al.} (1996) reported the
discovery of three new spectroscopically-confirmed LMC R CrB stars. These
stars reveal that the mean absolute magnitude for R CrB stars is likely
about one magnitude fainter than traditionally assumed. At present, about
one dozen R CrB stars have been discovered by the MACHO Project.

\item {\sl W Virginis/RV Tauri Stars} --- Alcock {\it et al.} (1997b) have
presented photometry and analysis for LMC W Virginis and RV Tauri variables.
Such stars have been relatively neglected because their longish-periods and
slightly erractic behavior makes short time-series difficult to interpret.
The period-luminosity (P-L) relation for these stars are consistent with
a common origin with the RV Tauri stars forming the long-period extension.
Futhermore, irregularities in the lightcurve appear first in the 15-20 {\sl day}
period range and get progressively larger with increasing luminosity (and
period).

\item {\sl HV 5756 - An Eclipsing Type II Cepheid} --- Welch {\it et al.} (1996)
report the discovery of the first type II Cepheid in an eclipsing binary
system. The Cepheid, HV 5756, was known from Harvard College Observatory
studies. However, the MACHO Project data reveals that this 17.5 {\sl day}
pulsator also undergoes eclipses at 419 {\sl day} intervals, the most recent
primary eclipse being 1997 Sep 11. The companion is a hotter star which
apparently has a luminosity significantly above that of the horizontal branch
in the LMC. This system provides the first opportunity to obtain a dynamical
mass for a type II Cepheid.

\item {\sl Long-period Variables} Cook {\it et al.} (1997) revealed the
presence of at least three sequences of long-period variables in a plot
of the surface brightness-corrected magnitude versus $\log$ P plot of singly-periodic
variables in the first-year LMC variable star analysis. These appear to
represent the fundamental, first-, and second-overtone sequences for LPVs.
The photometric amplitude of the lightcurves is strongly correlated with
the sequence.

\end{itemize}

\subsection{EROS}
\begin{center}
{\bf http://www.lal.in2p3.fr/EROS/presa.html}
\end{center}
The French microlensing survey, EROS, has produced a number of significant
results in the area of pulsating variables. Results reported in the literature
to date are from EROS 1 which employed a 0.6m telescope and produced very
extensive time-series (2-3$\times$ larger by epoch number) than MACHO, for
single fields in each of the LMC and SMC. At present an improved, larger-area
survey called EROS 2 is underway. The larger field and greater sensitivity
of the second project promise many new results for the study of variable stars
in the LMC and SMC. Major results related to pulsating variable stars are:

\begin{itemize}

\item {Cepheid P-L-C-[Fe/H]} --- Sasselov {\it et al.} (1997) have
used two-color photometry for 481 LMC and SMC Cepheids as well as sensible
reddening law constraints to derive a metallicity dependence for Cepheid
luminosities. They apply this to recent Hubble Space Telescope Cepheid survey
results and reinterpret the derived Hubble constant.

\item {SMC Beat Cepheids} --- Beaulieu {\it et al.} (1997) reported the
discovery of the first 11 beat Cepheids in the SMC. Four were fundamental/
first-overtone pulsators and the remainder were first/second-overtone
variables. The fundamental/first-overtone beat Cepheids have period ratios
higher than their LMC counterparts due to the lower metallicity of the
SMC population, as expected.

\item {LMC Classical Cepheids} --- Beaulieu {\it et al.} (1995) presented
the analysis of 97 Cepheid stars in the bar of the LMC from EROS 1 data.
They found one double-mode Cepheid and reported the clear separation of
first-overtone and fundamental mode sequences in the P-L relation.
Furthermore, they confirmed the existence of the long-period overtone
variables, sometimes called "long-period s-Cepheids".

\end{itemize}

\newpage

\subsection{OGLE}
\begin{center}
{\bf http://www.astrouw.edu.pl/~ftp/ogle/index.html}
\end{center}
The third large microlensing project, OGLE, concentrated on studying
the Milky Way bulge in its first incarnation. Largely confined to
observations in a single bandpass, and with the variable extinction
present along the line-of-sight, its pulsating star results towards
the bulge have been less useful than other surveys. However, it has
since studied variables in globular clusters and dwarf galaxies and
these works will likely be of great value in understanding the population
characteristics of low-mass pulsating stars.

The successor of the original project is OGLE 2, for which a new 1.3m 
telescope was built at Las Campanas, Chile. 

\section{HIPPARCOS}
The HIPPARCOS satellite mission and its firsts results indicate that variable
star populations studies have entered a new era. The most recent results can
be obtained from the European Space Agency's HIPPARCOS Web page:
\begin{center}
{\bf http://astro.estec.esa.nl/Hipparcos/hipparcos.html}
\end{center}
Specifically, the mission provides astrophysicists with three major databases:
1) multiple-epoch, uniform photometry, 2) high-precision proper motions, and
3) uniform, precise parallaxes. The very first results from the mission concentrated
on the Cepheid (Feast \& Catchpole, 1997) and RR Lyrae distance indicators - the
latter through their luminosity calibration by sub-dwarf parallaxes and the
subsequent re-determination of globular cluster distances (Reid, 1997).
Cepheids are sufficiently rare in the local volume that the parallax of only one 
star was significant. The interpretation of the Cepheid result is
further compunded by the fact that Polaris ($\alpha$ UMi) is an amplitude-changing
star whose pulsation mode can only be decided by consistency arguments.

The long-awaited parallaxes of nearby $\delta$ Scuti stars by Hog \& Petersen
(1997) confirm that the currently held belief that large-amplitude $\delta$ Scuti
stars are normal stars evolving through the instability strip. Another important
result is the first set of parallaxes from HIPPARCOS for 16 Mira (large-amplitude)
long-period variables by van Leeuwen {\it et al.} (1997). This study confirms
the presence of both fundamental-mode and first-overtone mode pulsators among
these variables, with the first-overtone mode being more common among the stars
studied.

\section{Algorithms}
\subsection{Robust Detection}

A serious consideration for large surveys is the number of ``false
positive''. Simply put, the fraction of stars which are true variables
in the field is roughly 1 in 250. However, the total number of stars
in a survey is typically 10$^7$. So a rate of false positives of 10$^{-4}$
would be required for the false positive sample to be the same order of
magnitude in size as the actual variable star sample. Clearly, a suppression
of false positives of a further factor of at least 100 would be desirable.

A workable and very robust technique for detecting variables in the
presence of ``photometric blunders'' was described by Welch \& Stetson
(1993). It makes use of the estimated uncertainty of each point and
the temporal coherence of the lightcurve to retain true variables
while rejecting the vast majority of false positives. Furthermore, 
straightforward tests for the presence of systematic errors are easily
implemented so that confidence regions for true variability with real
data are easily established.

\subsection{Automated Classification}

It has long been desirable to have a method of variable star 
classification which does not involve human intervention. Despite
this longing, until recently, even the variable stars discovered
in Hubble Space Telescope surveys were classified by eye. While the
relatively small number of stars in such surveys makes this a 
workable scheme, the number of variables generated by the large
microlensing surveys begs for something more automatic.

The first significant paper regarding automated classification of 
variables in the modern era is that of Stetson (1996). An algorithm 
is both described and tested in this remarkable paper which reduces 
unanalysed IC 4182 data from HST and compares it to a `classical' 
search made with earlier HST data. The automated technique produced 
essentially identical results for the Cepheids. An improved 
Lafler-Kinman statistic also is derived which reduces the effective 
weight of phase gaps. This paper should be required reading for anyone 
embarking on a large photometry program!

\subsection{Time-series Analysis}

A number of additional time-series analysis algorithms and tools have
become available in the last few years. In my mind, the most significant
of these is the "CLEANest" algorithm by Grant Foster, described in both
Foster (1995, 1996a, 1996b). This algorithm is {\it extremely} robust
for identifying real frequencies in the presence of very strong aliasing -
a problem which pervades astronomical time-series analysis. Starting from
the "Date-Compensated Discrete Fourier Transform", or DCDFT as it is more
commonly known, CLEANest is used to build a model of lightcurve behavior
which simultaneously incorporates all statistically-significant frequencies
and removes the effect of their aliases. 

A DOS version of this program with a graphical user interface written by
Foster which includes sample data files and documentation is available 
from the American Association of Variable Star Observers (AAVSO) site at
the URL:
\begin{center}
{\bf http://www.aavso.org/programs/software.html}. 
\end{center}
A Fortran77 version, coded by Andrew Rorabeck while at McMaster University, 
is available from the author. This version has no graphical interface and 
makes use of the commonly available ``Numerical Recipes'' subroutines of 
Press {\it et al.} (1986). Requests for electronic versions of the code may 
be made to {\sl welch@physics.mcmaster.ca}.

Another algorithm worthy of note is the ``Analysis of Variance'' or ``AoV''
method of Schwarzenberg-Czerny (1989) which is computationally far less
expensive than CLEANest.

\subsection{``Flux Difference'' Photometry}

Perhaps the most elegant new technique described in recent memory is ``flux
difference'' photometry. This is a procedure which allows variable objects
to be identified and studied in relatively high signal-to-noise data which
is confusion-limited. Said differently, in situations where several stars
may be within the angular bounds of a given pixel and hence any particular
one may be indistinguishable in an image, it is still possible to study the
nature of a variable object, so long as the zeropoint of the flux for a
single star is not required.

The most extensive description of the process of reducing data in this manner
is the paper by Tomaney \& Crotts (1996). This work, made significantly more
difficult than expected by a quickly varying point spread function across
the frames used for test purposes, reveals the high significance with which
strongly crowded variables may be detected. Immediate applications for the
search for microlensing using nearby galaxies are obvious. However, there
are numerous variable star populations programmes which might be most
profitably attacked with these methods. For instance, the period ratios of
double-mode stars may be measured as a function of galactocentric radius
in galaxies with metallicity gradients, such as M31. Studies of the time-series
of relatively rare stars of intermediate luminosity will also benefit by the 
larger effective search field that this technique employs.

\section{Conclusion}

A final word. It is my opinion that the most tasty fruit is yet to be picked
from the vine. It would be true to say that most of the published results
from the major surveys were the most easily extracted and understood. The
very long time-series produced by on-going surveys are ideal for studying
more difficult multimode behavior and secular trends in stellar properties. 
Ultimately, I believe that these will have the most lasting impact on progress 
in our understaning of pulsating stars. Clearly, much remains to be done and
the public availability of the time-series will ensure that the impact of
each project will continue to be felt long after the last exposure is taken
on a given telescope.

\acknowledgments

I acknowledge the support of a Natural Sciences and Engineering Research
Council of Canada (NSERC) Research Grant as well as the members of the
MACHO Project who have worked so closely with me over the past several
years and helped make my involvement with the collaboration so rewarding.

\medskip
The slides for the PowerPoint presentation given at the meeting are
available at the URL:
\begin{center}
{\bf http://wwwmacho.mcmaster.ca/LANL97/}
\end{center}

\end{document}